\documentclass[aps,pra,twocolumn,preprintnumbers,showpacs]{revtex4-1}
\usepackage{amsmath,amssymb,mathrsfs}
\usepackage{graphicx}
\usepackage{hyperref}
\usepackage[dvips]{color}

\def\tbf{\textbf}

\begin{document}

\title{Stripe, checkerboard, and liquid-crystal ordering from anisotropic $p$-orbital Fermi surfaces in optical lattices}

\author{Zixu Zhang}
\affiliation{Department of Physics and Astronomy, University of Pittsburgh, Pittsburgh, Pennsylvania 15260, USA}
\affiliation{Kavli Institute for Theoretical Physics, University of California, Santa Barbara, California 93106, USA}
\author{Xiaopeng Li}
\affiliation{Department of Physics and Astronomy, University of Pittsburgh, Pittsburgh, Pennsylvania 15260, USA}
\affiliation{Kavli Institute for Theoretical Physics, University of California, Santa Barbara, California 93106, USA}
\author{W. Vincent Liu}
\affiliation{Department of Physics and Astronomy, University of Pittsburgh, Pittsburgh, Pennsylvania 15260, USA}
\affiliation{Kavli Institute for Theoretical Physics, University of California, Santa Barbara, California 93106, USA}
\date{\today}

\begin{abstract}
We study instabilities of single-species fermionic atoms in the $p$-orbital bands in two-dimensional optical
lattices at noninteger filling against interactions. Charge density wave and orbital density wave
orders with stripe or checkerboard patterns are found for attractive and repulsive interactions, respectively. The superfluid phase,
usually expected of attractively interacting fermions, is strongly suppressed. We also use field theory to analyze the possible phase-transitions from orbital stripe order to liquid-crystal phases and obtain the phase diagram. The condition of nearly-perfect Fermi-surface nesting, which is key to the above results, is shown robustly independent of fermion fillings in such $p$-orbital systems, and the $(2 k_F ,\pm 2k_F)$ momentum of density wave oscillation is highly tunable. Such remarkable features show the promise of making those exotic orbital phases, which are of broad interest in condensed-matter physics, experimentally realizable with optical lattice gases.
\end{abstract}
\preprint{NSF-KITP-11-074}


\maketitle
\section{introduction}
Ultracold atoms in optical lattices are highly tunable quantum systems, which
are ideal to simulate conventional condensed-matter physics. Traditional
studies carried out on optical lattices only involve the lowest $s$ band.
In recent years, studies on higher orbital bands of optical lattices have shown many
interesting results~\cite{orbitaldance}. For bosons, staggered orbital orders in square lattices
\cite{PhysRevA.72.053604,PhysRevA.74.013607,PhysRevLett.97.110405,PhysRevLett.108.175302} and stripe orders in triangle lattices
\cite{PhysRevLett.97.190406} have been proposed. Most recently, experimental
progress has realized superfluidity on higher-orbital bands
\cite{PhysRevLett.99.200405, hemmerich1,Soltan-Panahi2012}, and experimental signatures distinguishing the staggered orbital order have been proposed in recent theoretical work~\cite{PhysRevA.83.063626,2011arXiv1110.3021C,PhysRevLett.108.175302}. For fermions, frustrated
orbital orders in the strongly interacting Mott regime were found
\cite{PhysRevLett.100.160403, PhysRevLett.100.200406} years ago. Recently, novel Fulde-Ferrell-Larkin-Ovchinnikov (FFLO) phases
\cite{PhysRev.135.A550, fflo2} on the $p$-orbital bands
\cite{PhysRevB.83.144506, PhysRevA.83.063621} and multiband superconductivity \cite{PhysRevA.82.033610} in optical lattices have been proposed, keeping this field
fascinating.

In the present work, we study the interacting spinless $p$-orbital
fermionic atoms in two-dimensional (2D) square optical lattices with both attractive and  repulsive interactions. We find that the quasi-one-dimensional feature of the Fermi surfaces of the double degenerate $p_x$- and
$p_y$-orbital bands gives rise to the following interesting orders. For attractive interactions, it induces charge density wave (CDW) order in a wide filling regime where the superfluid order is greatly suppressed. For repulsive interactions, orbital density wave (ODW) order is induced. Both CDW and ODW show stripe or checkerboard patterns in space, depending on the filling. We further show that our system is a simple, clean, and highly tunable system to realize possible nematic and smectic liquid-crystal phases, which is a topic of great current interest in correlated condensed-matter physics \cite{PhysRevB.64.195109,benjaminNJP,chungweilin,2011arXiv1112.0773R}.

\section{System and model}

Consider a system of spinless fermions filled up to degenerate
$p_x$- and $p_y$- orbital bands in a 2D square lattice. Such a system can be realized by considering an anisotropic three-dimensional optical lattice with lattice potential $V_{\textrm{op}}=\sum_{\nu=x,y,z}V_{\nu}\sin^2(k_L r_{\nu})$, where $k_L$ is the wave vector of the laser beams and the lattice constant is $a=\pi/k_L$. By setting $V_z\gg V_x=V_y$, we realize dynamically decoupled 2D square-lattice layers in the $xy$ plane, each being a 2D system. The 2D system is then filled with spinless fermions such that the
lowest $s$ band is fully occupied and two degenerate $p_x$- and $p_y$- orbital bands are partially filled. In general, the band gap between the $s$
and $p$ bands is much larger than the interaction, and the $s$-band
fermions are dynamically inert. By expanding the fermionic field operators in the Wannier basis and using a tight-binding approximation, we obtain the $p$-band Fermi-Hubbard model
\begin{eqnarray}
H&=&\sum_{\mathbf{r}\alpha \beta} t_{\alpha \beta}
(C^{\dagger}_{\alpha,\mathbf{r}+\mathbf{e_{\beta}}}
C_{\alpha,\mathbf{r}}+h.c.)\nonumber\\
&& -\mu \sum_{\mathbf{r} \alpha }n_{\alpha,\mathbf{r}}
+g\sum_{\mathbf{r}}n_{x,\mathbf{r}}n_{y,\mathbf{r}}
\label{eq:ham1}
\end{eqnarray}
to describe the system with chemical potential $\mu$. Equation~\eqref{eq:ham1} only contains nearest-neighbor hopping and onsite interaction, since in typical ultracold-atom experiments the next-nearest-neighbor hopping and nearest-neighbor interaction are negligible. In Equation~\eqref{eq:ham1}, $C_{\alpha,\mathbf{r}}$ is the annihilation operator of Wannier state
$p_{\alpha}$ at site $\mathbf{r}$, and
$n_{\alpha,\mathbf{r}}=C_{\alpha,\mathbf{r}}^{\dagger}C_{\alpha,\mathbf{r}}$
is the number operator for the $p_{\alpha}$-orbital state at site
$\mathbf{r}$. The subscripts $\alpha$ and $\beta$ run over $x$ and $y$. The hopping term $t_{\alpha \beta}$ is given by $t_{\alpha \beta}=\left[t_{\parallel}\delta_{\alpha
\beta}-t_{\perp}(1-\delta_{\alpha\beta})\right]$, where the parallel (transverse) hopping $t_\parallel$ ($t_\perp$) means the hopping of $p_\alpha$-orbital fermions at site $\mathbf{r}$ to the nearest neighbor $\mathbf{r}+\mathbf{e}_\beta$, with $\beta=\alpha$ ($\beta\neq\alpha$). Here, $\mathbf{e}_\alpha$ is the lattice unit vector in the $\alpha$ direction. The last term is
the onsite interaction between $p_x$- and $p_y$-orbital fermions induced
by $p$-wave scattering, with $g$ as the coupling constant. Due to the $p$-orbital anisotropy, we expect $t_\parallel \gg t_\perp$. In quantum chemistry, $t_\parallel$ ($t_\perp$) is referred to as the $\sigma$ ($\pi$) bond. We can use a harmonic approximation to estimate $t_\parallel$ and $t_\perp$, where a standard tight-binding calculation gives the transverse hopping $t_{\perp}=e^{-(\eta/2)^2}V_x/2$, and the parallel hopping $t_{\parallel}=|\eta^2/2-1|t_{\perp}$. The parameter $\eta = \alpha_x a$ is typically a large number ($\gg 1$) and therefore $t_{\parallel} \gg t_\perp$.  Here, $\alpha_{\nu}=(V_{\nu}/E_R)^{1/4}k_L$, where $E_R=\hbar^2 k_L^2/2m$ is the recoil energy for atoms of mass $m$. The onsite interaction is given by $g=g_p \alpha_x^2 \alpha_z (22\alpha_x^2+\alpha_z^2)/32(2\pi)^{3/2}$ in the pseudopotential approach with coupling constant $g_p$~\cite{PhysRevLett.100.160403}.  In the paper, the lattice constant, Boltzmann constant, and Planck constant are all set to be one.

\section{Fermi surface instabilities}
\label{sec:fsi}
\begin{figure}[t]
\includegraphics[width=0.6\linewidth]{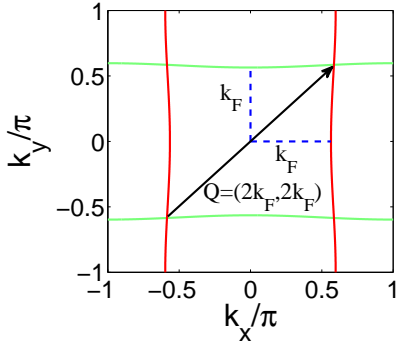}
\caption{(Color online) A schematic diagram illustrating how the $(2k_F, 2k_F)$ momentum
of density fluctuation satisfies the nesting Fermi-surface condition. Here the lattice constant $a$ is set to be one. Red (dark gray) solid
curve: Fermi surfaces of $p_x$-orbital band. Green (light gray) solid
curve: Fermi surfaces of $p_y$-orbital band. Blue dashed line: Fermi momenta of $p_x$- and $p_y$-
orbital bands. Solid arrow: the $(2k_F, 2k_F)$ momentum of density fluctuation simultaneously satisfying the nesting Fermi-surface condition for both
$p_x$- and $p_y$- orbital bands.}
\label{fig:cartoon}
\end{figure}
It is well known that nesting Fermi surfaces are crucial
to realize some
spontaneously translational-symmetry-breaking phases, e.g., CDW, spin density wave~(SDW),
and FFLO. For the lowest $s$ band in a 2D square lattice, the nesting Fermi
surfaces only occur
at half filling, assuming only nearest-neighbor hopping. In contrast, in our system, the nesting of quasi-one-dimensional $p_x$ and $p_y$ Fermi surfaces, as shown in Figure~\ref{fig:cartoon} is independent of filling for a wide range of $\mu$, as long as
$t_{\perp} \ll t_{\parallel}$. In Figure~\ref{fig:cartoon}, $p_x$ and $p_y$ Fermi surfaces are perpendicular to each other, which greatly suppresses the Cooper instability from particle-particle channel scattering. The reason is that in order to induce Cooper instability, all of the fermion pairs need to have almost the same center-of-mass
momentum, which is impossible here with each particle-particle pair composed of one $p_x$- and one $p_y$-orbital fermion, given only onsite interaction in Equation~\eqref{eq:ham1}. In contrast, each $p_x$ ($p_y$) particle-hole pair in the density channel is composed of one particle and one hole within the $p_x$- ($p_y$-) orbital band, which benefits from the nesting Fermi-surface condition. To
simultaneously satisfy the nesting Fermi-surface condition for both $p_x$-
and
$p_y$- orbital bands in the density channel, the momentum of density fluctuation should be
\begin{equation}
\textstyle{\mathbf{Q_{1,2}} \approx(2k_F, \pm 2k_F),}
\label{eq:momentum}
\end{equation}
as shown by the black arrow in
Figure~\ref{fig:cartoon}, where $k_F$ is the Fermi momentum for each band.

To illustrate the above statement quantitatively, we study different instabilities in our system by random-phase approximation (RPA) as follows. The validity of RPA is discussed in Appendix~\ref{sec:RPAapp}. For the density channel, we define the density operator $\rho_{\alpha,q}=\sum_{k}C_{\alpha,k+q}^{\dagger}C_{\alpha,k}$ in momentum-Matsubara frequency
space for the $p_\alpha$- orbital band where $\alpha=x,y$. The 2+1 momenta $k$ and $q$ are defined as $k \equiv (\mathbf{k},i\omega_m)$ and $q
\equiv (\mathbf{q},i\omega_n)$, where $\omega_m=(2m+1)\pi T$ and
$\omega_n=2n\pi T$ are fermionic and bosonic Matsubara
frequencies. The density-density
correlation function in our system without interaction has the form
\begin{eqnarray}
\Pi^0_{\alpha \beta}(q)&=&\frac{T}{N^2} \langle \rho_{\alpha,q}
\rho_{\beta, -q} \rangle^0\nonumber\\
&=&-\frac{\delta_{\alpha \beta}}{N^2}
\sum_{\mathbf{k}}\frac{n_F(\xi_{\alpha,\mathbf{k}})-n_F( \xi_{\alpha,
\mathbf{k+q}})}{i\omega_n-\xi_{\alpha, \mathbf{k+q}}+\xi_{\alpha,
\mathbf{k}}},
\label{eq:cdwsdwdiv0}
\end{eqnarray}
where $\langle ... \rangle^0$ means thermal average without interaction. The spectrum of $p_x$-orbital fermions is $\xi_{x, \mathbf{k}}=2t_{\parallel}\cos k_x -2t_{\perp} \cos k_y -\mu$, and $\xi_{y, \mathbf{k}}$ has a
similar form. The Fermi distribution function $n_F$ is given by $n_F(\xi_k)=1/(e^{\xi_k/T}+1)$. At $\mathbf{q}=\mathbf{Q}=(2k_F,\pm 2k_F)$, ${t_\perp \rightarrow 0}$ (perfect nesting) and
$i\omega_n=0$
(static limit), Equation~\eqref{eq:cdwsdwdiv0} reduces to
\begin{equation}
\chi^0 \equiv \Pi^0_{xx}=\Pi^0_{yy} \sim D(E_F) \textrm{ln}\left(
\frac{\omega_D}{k_B T}\right)
\label{eq:cdwsdwdiv}
\end{equation}
in the continuous limit. In Equation~\eqref{eq:cdwsdwdiv},
$D(E_F)$ is the density of states near the Fermi surface and
$\omega_D$ is some energy cutoff.

\begin{figure}[t]
\includegraphics[width=0.85\linewidth]{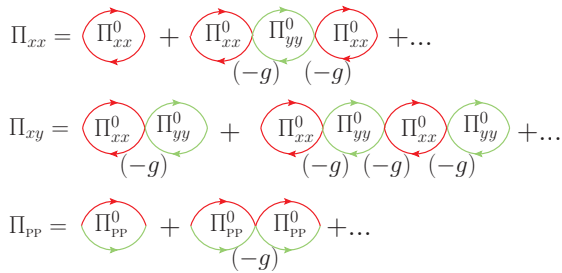}
\caption{(Color online) Feynman diagrams of density-density correlation functions of
the same orbital band $\Pi_{xx}$ and different orbital bands $\Pi_{xy}$, and the pair-pair
correlation function $\Pi_{_\textrm{PP}}$. Red (dark gray) and green (light gray) lines are the propagators of free $p_x$ and $p_y$ fermions.}
\label{fig:channels}
\end{figure}

When the interaction is turned on, the
density-density correlation function $
\Pi_{\alpha \beta}(q)=\frac{T}{N^2} \langle \rho_{\alpha,q}
\rho_{\beta, -q} \rangle
$
can be evaluated by RPA, as shown in Figure~\ref{fig:channels}, where $\langle ... \rangle$ means
thermal average with interaction. We then evaluate the following
correlation functions:
\begin{equation}
\Pi_{\pm}=\frac{T}{N^2}\langle \rho_{\pm}(q)\rho_{\pm}(-q)\rangle,
\label{eq:cdwodw}
\end{equation}
where $\rho_{\pm}(q)=\rho_{x,q}\pm\rho_{y,q}$
are the total density and density difference between $p_x$ and $p_y$ fermions, {i.e., the CDW and ODW instability channels}. Given that $t_\perp \rightarrow 0$,
$\mathbf{q}=\mathbf{Q}=(2k_F, \pm 2k_F)$, and $i\omega_n=0$, Equation~\eqref{eq:cdwodw} reduces to
\begin{equation}
\Pi_{\pm}=
\frac{2\chi^0}{1\pm g\chi^0}.
\label{eq:finalins}
\end{equation}
Since $\chi^0 \sim D\textrm{ln}\left(\frac{\omega_D}{T}\right)$, any arbitrarily small attractive
(repulsive) interaction $g<0$ ($g>0$) can induce divergence of $\Pi_+$~($\Pi_-$) at sufficiently low temperature. Such divergence indicates phase-transition to the corresponding
symmetry-breaking phase.

In experiments, a small but finite $t_\perp$ is inevitable, which makes the Fermi-surface nesting not perfect. The density wave orders do not exist even at $T=0$ if the interaction is too small. However, for a nonperfect nesting, the same density wave orders are generally expected to exist if the interaction strength exceeds a certain critical value. For example, even if the Fermi surfaces between spin-up and -down fermions are Zeeman split in the presence of a field, it is well known that the BCS superfluidity or superconductivity persists up to a critical Zeeman splitting for a given interaction strength. The latter is known as the Chandrasekhar-Clogston limit~\cite{cclimit1, cclimit2}. Increasing $|g|$ is experimentally feasible due to the high tunability of the interaction in optical lattices, e.g., by increasing the lattice potential.

\begin{figure}[t]
\includegraphics[width=0.9\linewidth]{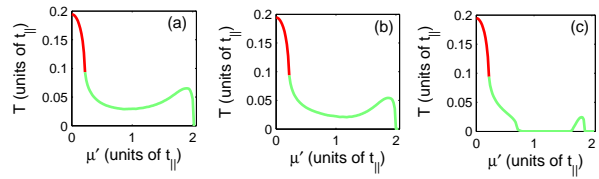}
\caption{(Color online) phase-transition temperature for CDW instability at $g=-2$ from RPA calculation. $t_\perp=$ (a) 0, (b) 0.04, (c) 0.08. Red (dark gray) line: instability towards the checkerboard density wave order. Green (light gray) line: instability towards stripe density wave order. The transition temperature towards checkerboard density wave order with the effective chemical potential (including Hartree term) $\mu'$ near $0$ is much higher than that towards the stripe density wave order with $\mu'$ away from $0$, which indicates that the former is much stronger than the latter. This feature comes from the Umklapp process at half filling. Besides, the phase-transition temperature towards the checkerboard density wave order does not show any noticeable change as $t_\perp$ increases from $0$ to $0.08$, which suggests that the checkerboard density wave order is not affected by $t_\perp$.}
\label{fig:pdrpa}
\end{figure}

Figure~\ref{fig:pdrpa} shows the phase-transition temperatures from CDW instability evaluated by RPA with small transverse hoppings, $t_\perp=0, 0.04$, and $0.08$. We set the system size to be $N^2=300^2$. The parallel hopping is set to be $t_\parallel=1$ as the energy unit, and we choose the interaction strength $g=-2$. It can be seen that although a small $t_\perp=0.04$ weakens the (stripe) density wave order, at a finite interaction $g=-2$, the (stripe) density wave order still occurs. However, a stronger $t_\perp=0.08$ destroys the (stripe) density wave orders over a certain range of the effective chemical potential (including Hartree term) $\mu'$.

The particle-particle (Cooper) channel can be studied in a similar way by evaluating the correlation function of the pair operator, $\Delta_{q}= \sum_{k} C_{x,-k+q} C_{y, k}$.
The pair-pair correlation function in our system without interaction reads
\begin{eqnarray}
\Pi^0_{_\textrm{PP}}(q)&=&\frac{T}{N^2}\langle \Delta^{\dagger}_q
\Delta_q\rangle^0,\nonumber\\
&=&-\frac{1}{N^2}\sum_{\mathbf{k}}
\frac{1-n_F(\xi_{x,\mathbf{-k+q}})-n_F(\xi_{y,\mathbf{k}})}
{i\omega_n-\xi_{x,\mathbf{-k+q}}-\xi_{y,\mathbf{k}}}.
\label{eq:fflochi}
\end{eqnarray}
Recall that for a density-density correlation function without interaction, by choosing $\mathbf{q}=\mathbf{Q}=(2k_F,\pm 2k_F)$, Equation~\eqref{eq:cdwsdwdiv0} has logarithmic divergence as shown in Equation~\eqref{eq:cdwsdwdiv}. In contrast, no logarithmic divergence is found in Equation~\eqref{eq:fflochi} at any value of $\mathbf{q}$. Therefore, the pair-pair correlation with interaction,
\begin{equation}
\Pi_{_\textrm{PP}}=\frac{T}{N^2}\langle \Delta^{\dagger}_q \Delta_q
\rangle=\frac{\Pi_{_\textrm{PP}}^0}{1+g \Pi_{_\textrm{PP}}^0}
\end{equation}
evaluated by RPA as shown in Figure~\ref{fig:channels},
will not diverge at any temperature, which is different from the density-density correlation functions in Equation~\eqref{eq:finalins}. It means that the instability of the
particle-particle
channel is greatly suppressed, and there is no phase-transition towards superconductivity.

\section{Mean field theory at $T=0$}
\begin{figure}[t]
\includegraphics[width=0.83\linewidth]{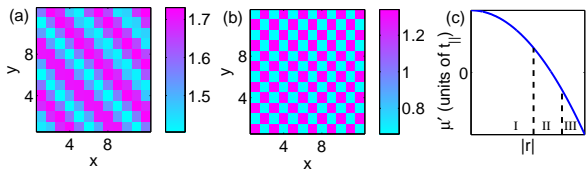}
\caption{(Color online)  The spatial total density pattern of CDW obtained from real space mean-field analysis (showing $12^2$ out of $300^2$ sites) with $g=-2$ at (a) $\mu=0$ and (b) $\mu=-1$. For ODW ($g=2$), the density difference shows similar patterns to (a) and (b) for $\mu=0$ and $\mu=1$, respectively. (c) A schematic LDA phase diagram in the presence of a trap.  $|\mathbf{r}|$ is the distance from the center of the trap. I, II, and III are the regions of stripe, checkerboard, and stripe density wave orders, respectively. Here $x$, $y$, and $\mathbf{r}$ represent the site numbers.}
\label{fig:cdwodw}
\end{figure}

The above consideration only shows that a phase-transition towards density wave order can happen in our system. In order to find the ground-state property, i.e., the order parameter, we apply a real-space mean-field analysis at $T=0$ for both $g>0$ and $g<0$. The
interaction part of the Hamiltonian given by Equation~\eqref{eq:ham1} can be decoupled in the density channel such that
\begin{equation}
\sum_{\mathbf{r}}n_{x,\mathbf{r}}n_{y,\mathbf{r}} \approx \sum_{\mathbf{r}}
(n_{x,\mathbf{r}}M_{y,\mathbf{r}}+
n_{y,\mathbf{r}}M_{x,\mathbf{r}}-M_{x,\mathbf{r}}M_{y,\mathbf{r}}),
\label{eq:cdwdecouple1}
\end{equation}
where $
\langle n_{\alpha,\mathbf{r}} \rangle=M_{\alpha,\mathbf{r}}
$
is the self-consistent condition and $\langle ... \rangle$ means the expectation value of the ground state at $T=0$. Eq.~\eqref{eq:ham1} reduces to
\begin{eqnarray}
H_{_{\textrm{MF}}}&=&\sum_{\mathbf{r}\alpha \beta}t_{\alpha \beta}
(C^{\dagger}_{\alpha,\mathbf{r}+\mathbf{e_{\beta}}}
C_{\alpha,\mathbf{r}}+h.c.)-\mu \sum_{\mathbf{r} \alpha }n_{\alpha,\mathbf{r}}\nonumber\\
&&+g \sum_{\mathbf{r}} (n_{x,\mathbf{r}}M_{y,\mathbf{r}}+
n_{y,\mathbf{r}}M_{x,\mathbf{r}}-M_{x,\mathbf{r}}M_{y,\mathbf{r}}),\nonumber\\
\label{eq:bdg}
\end{eqnarray}
which is in quadratic form and can be solved self-consistently. We set the parameters the same as before in Sec.~\ref{sec:fsi}, with $t_{\perp}=0$ to simplify the calculation. For attractive interaction $g=-2$, we find CDW order where the densities of $p_x$ and $p_y$ fermions are the same. When $\mu=0$, the total density of the ground
state exhibits a stripe pattern in real space as shown in Figure~\ref{fig:cdwodw}(a), and the energy per site is $-2.0295$, lower than that of the homogeneous-density state of $-2.0282$. When $\mu=-1$, the total density exhibits
a checkerboard pattern as shown in Figure~\ref{fig:cdwodw}(b), and the system is at half filling. In this case, the ground-state energy per
site is $-0.7826$, which is lower than that of the homogeneous-density state of $-0.7731$. Fourier series $M_{\alpha,\mathbf{r}}=a_0+\sum_{n=1}^{\infty} \left[ a_n\cos (n\mathbf{ q
\cdot r}) + b_n\sin (n\mathbf{ q \cdot r}) \right]$ are then used to fit Figure~\ref{fig:cdwodw}(a) and (b). It can be seen that the chemical potential is modified to $\mu'= \mu-a_0g$
by the background density $a_0$ (Hartree term), and the filling is determined by
$\mu'$ instead of $\mu$. We find $\mathbf{q} \approx (0.42 \pi, 0.42\pi)$ by fitting Figure~\ref{fig:cdwodw}(a). Higher-order harmonics (the $n>1$ Fourier components) are found to be nonvanishing in this case as expected in CDW, but are very weak compared to the first-order terms ($a_n \ll a_1$, $b_n\ll b_1$, $n>1$). The checkerboard pattern in Figure~\ref{fig:cdwodw}(b) has momentum $\mathbf{q}=(\pi,\pi)$. The $\mathbf{q}$'s in both cases agree with Equation~\eqref{eq:momentum} very well, with $k_F$ determined by the effective chemical potential $\mu'$.

For repulsive interaction $g=2$, we find ODW order, where the
densities of $p_x$- and $p_y$-orbital fermions are no longer the same, and the difference between them (ODW order parameter) oscillates in space. When $\mu=0$ and $\mu=1$, the density difference shows stripe and checkerboard patterns similar to Figure~\ref{fig:cdwodw}(a) and (b).

There are some general properties of the density wave orders. (1) The checkerboard order at half filling is much stronger than the stripe order with other fillings due to the Umklapp process, which greatly enhances density wave order at half filling \cite{RevModPhys.66.129}. (2) In general, increasing transverse hopping $t_\perp$ weakens the nesting Fermi surface condition and tends to destroy the
stripe order.  However, the checkerboard order at half filling is not affected by the Fermi-surface curvature. It is because a
$(\pi,\pi)$ momentum of checkerboard order always satisfies the perfect nesting condition, independent of $t_{\perp}$~\cite{RevModPhys.66.129}. The above two features can be studied by calculating instabilities as shown in Sec.~\ref{sec:fsi} or evaluating energy gain with mean-field analysis. (3) The Neel orbital
order found in Ref.~\cite{PhysRevLett.100.160403} at half filling and
strong-coupling limit $g\rightarrow +\infty$, where $p_x$ and $p_y$ Wannier orbitals alternate in space, can be understood as the extreme case of
checkerboard ODW.

In experiments, a shallow harmonic trap $V(\mathbf{r})$ is present in addition to the optical lattice, and a spatial phase separation is expected due to the additional trapping potential. With local density approximation~(LDA) $\mu(\mathbf{r}) \rightarrow \mu(\mathbf{r})-V(\mathbf{r})$, a schematic phase diagram is shown in Fig.~\ref{fig:cdwodw}(c). At the center region of the trap where the local chemical potential is the highest, the stripe order exists due to large filling (region I). As one moves towards the edge of the trap, the filling decreases, and when the effective local chemical potential $\mu'(\mathbf{r}) \approx 0$, the checkerboard order appears and this region is at half filling (region II). As one moves towards the edge further, the filling becomes low and the stripe order emerges again (region III). Therefore, our theory predicts a spatial density profile of phase separation with stripe core $\rightarrow$ checkerboard shell $\rightarrow$ stripe edge.

\section{liquid-crystal phases at $T\neq0$}
\label{sec:eft}
The above mean-field analysis shows the existence of density wave orders at $T=0$. As one raises the temperature, the thermal melting effect may drive the system to different liquid-crystal phases before it becomes normal Fermi liquid. In the following part, we will first present a field theory which incorporates the thermal melting effect to study the liquid-crystal phases in square-lattice systems. Then we will make connections between this field theory and the specific microscopic model discussed before, i.e., how to determine the coefficients in the field theory from Equation~\eqref{eq:ham1}. At last, we will comment on the advantages of our system to study liquid-crystal phases.

For simplicity, we only consider liquid-crystal phases from stripe CDW with attractive interaction, where the densities of $p_x$- and $p_y$-orbital fermions are the same. ODW with repulsive interactions are left for future study. The stripe CDW breaks the C$_4$ rotational symmetry of the square lattice down to $\textrm{C}_2$ (a Z$_2$ phase-transition), and also breaks lattice translational symmetry, as shown in Figure~\ref{fig:cdwodw}(a). As a result, two types of topological defects can occur at finite temperature: the Z$_2$ domain walls and the (edge) dislocations of stripes, where the latter may drive the system to smectic or nematic liquid-crystal phases. The smectic liquid-crystal phase breaks both translational symmetry and $\textrm{C}_4$ rotational symmetry, which is essentially the same as the stripe order, while the nematic liquid-crystal phase only breaks $\textrm{C}_4$ rotational symmetry and can be viewed as melted smectic stripes~\cite{knature, PhysRevB.78.085124}.

The total density fluctuations associated with momenta $\mathbf{Q_{1,2}}$ can be parameterized as $\delta \rho= \left[ \phi_1 e^{i\mathbf{Q_1 \cdot r}} + \phi_2 e^{i\mathbf{Q_2 \cdot r}} +c.c. \right]$. For the incommensurate case, by adopting Gaussian approximation and keeping only up to quartic terms, the effective action reads
\begin{eqnarray}
S&=&{\textstyle \frac{1}{T}\int d^2 r~\sum_{\sigma=1,2} \left(j|\nabla \phi
_\sigma|^2+r|\phi_\sigma|^2+u|\phi_\sigma|^4\right)}\nonumber\\
&&{\textstyle +v|\phi_1|^2|\phi_2|^2 + S_t + ...,}
\label{eq:lcgl1}
\end{eqnarray}
where $S_t$ denotes topological defects of stripe dislocations, similar to the vortex term in the  $XY$ model. Eq.~\eqref{eq:lcgl1} is invariant under C$_4$ rotations of $\pi/2$, $\pi$, and $3\pi/2$, which yields $(\phi_1, \phi_2) \rightarrow (\phi_2^*, \phi_1), (\phi_1^*, \phi_2^*)$, and $(\phi_2, \phi_1^*)$, respectively. Besides, Eq.~\eqref{eq:lcgl1} also has two U(1) symmetries, i.e., $\phi_{1,2} \rightarrow  \phi_{1,2} e^{i\varphi_{1,2}}$, where $\varphi_{1}$ and $\varphi_{2}$ are arbitrary global phases. The coupling constants $j,r,u,$ and $v$ can be derived from the microscopic model given by Equation~\eqref{eq:ham1} as shown in Appendix~\ref{sec:eftapp}, and we find $v=4u$, which strongly suppresses the coexistence of $\phi_1$ and $\phi_2$. Without loss of generality, we assume $\phi_2$ is suppressed, i.e., the saddle point is at $|\phi_1|=\Phi$ and $|\phi_2|=0$, and write $\phi_1=\Phi e^{i\varphi}$. Neglecting the (gapped) amplitude fluctuation of $\Phi$, the low-energy theory is described by $\varphi$ as $S_\varphi=\frac{1}{T}\int d^2 r~ j \Phi^2 (\nabla \varphi)^2$.

The smectic and nematic order parameters can be defined as $\langle \phi_1 -\phi_2 \rangle$ and $ \langle |\phi_1|^2 - |\phi_2|^2 \rangle$, respectively. At $T=0$, $\langle \phi_1 -\phi_2 \rangle \sim \langle \phi_1 \rangle \sim \Phi e^{i\varphi_0}\neq 0$, with $\varphi_0$ as an arbitrary global phase, and the system is smectic. At arbitrary small temperature, the gapless U(1) mode of $\varphi$ restores the translational symmetry, causing $\langle \phi_1 \rangle=\langle \phi_2 \rangle = 0$. The system is algebraic smectic, with algebraic order of $\phi_1$. As the temperature increases, the stiffness $J$ defined as $J/2 \equiv j \Phi^2/T$ decreases from infinity according to microscopic calculation.  When $J$ reaches $2/\pi$, the system undergoes a Kosterlitz-Thouless transition and the algebraic order of $\phi_1$ is destroyed by proliferation of stripe dislocations, similar to the destruction of superfluidity by vortices in the $XY$ model. The system becomes nematic with a short-range correlation of $\phi_1$, while the C$_4$ rotational symmetry remains broken. Further increasing of temperature eventually drives a second-order Ising-nematic phase-transition, above which the $\textrm{C}_4$ rotational symmetry is restored with $\langle |\phi_1|^2 \rangle=\langle|\phi_2|^2\rangle$, and the system becomes normal.

For the commensurate case with momentum $2k_F=2\pi p'/p$, where $p'$ and $p$ are relatively prime
integers, an additional term $w \cos(p \varphi)$ is allowed in Equation~\eqref{eq:lcgl1}. The U(1) symmetry of $\phi_1$ is reduced to Z$_p$ here, which means the action is invariant under translation $\varphi \rightarrow \varphi +2\pi p''/p$, with non-negative integer $p''<p$. With this additional cosine term, the theory of $\varphi$ naturally reduces to the Z$_p$ compact clock model. According to the renormalization-group analysis of the compact clock model \cite{footnote4}, our system undergoes the smectic$-$nematic$-$normal transition as one increases the temperature when $1<p \le 4$, and the smectic$-$algebraic smectic$-$nematic$-$normal transition when $p>4$. As $p \rightarrow \infty$ where the system approaches incommensurate, the smectic$-$algebraic smectic phase-transition temperature reduces to zero, which is consistent with the incommensurate case discussed before.

\begin{figure}[t]
\includegraphics[width=0.85\linewidth]{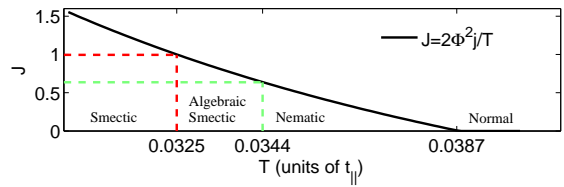}
\caption{(Color online) Phase diagram at $1/5$ filling where $p=5$. Red (dark gray) dashed line: determining the smectic$-$algebraic smectic phase-transition temperature at $J=p^2/8\pi$. Green (light gray) dashed line: determining the algebraic smectic$-$nematic phase-transition temperature at $J=2/\pi$.}
\label{fig:lcpd}
\end{figure}

As a specific, nontrivial example to make connections between the field theory given by Equation~\eqref{eq:lcgl1} and the microscopic model given by Equation~\eqref{eq:ham1}, a system with commensurate $1/5$ filling ($p=5$) is studied. We focus on the temperature regime near the Ising-nematic phase-transition point, so that in Equation~\eqref{eq:lcgl1} the approximation $\Phi^2=-r/2u$ for the saddle point is applicable. The procedure is to write Equation~\eqref{eq:ham1} in path-integral form and use Hubbard-Stratonovich transformation to introduce the density fields and decouple the interaction term. The fermionic fields are then integrated out and the remaining density fields reproduce the field theory given by Equation~\eqref{eq:lcgl1} with coefficients calculated. The details of the procedure are shown in Appendix~\ref{sec:eftapp}. After obtaining the coefficients $r,j,u,$ and $v$, the phase-transition temperature can be determined from the value of $J$ according to the previous discussion. The phase diagram is shown in Figure~\ref{fig:lcpd}.

Our system has the following advantages to study liquid-crystal phases. (1) There are no other competing orders. A similar spinful condensed-matter system was studied~\cite{PhysRevB.74.245126}, which showed that a stripe CDW order of momentum $(2k_F, 2k_F)$ is competing with a checkerboard CDW order of coexisting momenta $(2k_F,\pi)$ and $(\pi,2k_F)$ \cite{footnote2}. Also, a similar spinful ultracold-atomic system that has much more complicated interaction~\cite{PhysRevA.83.063621} may have many competing phases such as BCS, which is not as clean and simple as the proposed spinless system to study the liquid-crystal phases.  (2) The $2k_F$ momentum dependence of density wave order is highly tunable by changing the fillings, which makes it easy to adjust the commensurability.

\section{Experimental realization}
Our system can be realized by loading fermionic atoms such as $^{40}$K or $^6$Li of a single hyperfine state on square optical lattices. The fermions will automatically occupy $p$-orbital bands after the $s$ band is fully filled. The interactions between spinless fermions can be tuned by $p$-wave Feshbach resonance, together with controlling the lattice spacing and potential depth. The cold gas is not required to be so close to the resonance, so the atom loss rate can be kept relatively low. The momenta of density wave orders $\mathbf{Q_{1,2}}$ can be detected by
optical Bragg scattering~\cite{PhysRevLett.107.175302}.
Alternatively, {\it in situ} imaging can directly show the density pattern $\delta \rho$. As discussed before, at half filling the checkerboard density waves are greatly enhanced by the Umklapp process, and are not affected by the Fermi-surface curvature caused by transverse hopping. Therefore, in experiments one should first search for the checkerboard density waves at half filling, which has the phase-transition temperature $T_c \sim 0.2 t_\parallel$ based on RPA (a mean-field-level estimate) as shown in Sec.~\ref{sec:fsi}, given that $|g/t_\parallel|=2$. In addition, the hopping $t_\parallel$ of $p$-band fermions is in general an order of magnitude larger than the $s$-band hopping, which also enhances the phase-transition temperature. With typical experimental parameters such that $\lambda \sim 500$nm, $V_x \sim 5 E_R$, and using $^{40}$K atom, the estimated parallel hopping from harmonic approximation is $t_\parallel~\sim 100$nk, and then the phase-transition temperatures in Figure~\ref{fig:pdrpa} and~\ref{fig:lcpd} can be determined. For example, the estimated phase-transition temperature for the checkerboard density waves at half filling is $T_c \sim 20$~nk from Figure~\ref{fig:pdrpa}.

\section*{ACKNOWLEDGEMENTS}
We thank Kai Sun, Meng Cheng, Zi Cai, and Chungwei Lin for helpful discussions. This research is supported in part by ARO (Grant No. W911NF-11-1-0230),  DARPA-OLE-ARO (Grant No. W911NF-07-1-0464), and NSF (Grant No. PHY05-51164).

\appendix
\section{Validity of RPA}
\label{sec:RPAapp}
We justify the validity of RPA used in Sec.~\ref{sec:fsi} as follows.
Our system can be viewed as two sets of noninteracting one-dimensional (1D) spinless Fermi chains perpendicular to each other, with one set in the $x$-direction and the other in the $y$-direction. The $x$ chains ($y$ chains) are weakly coupled by small transverse interchain hopping in the $y$ ($x$) direction, and show quasi-one-dimensional Fermi surfaces (Figure~\ref{fig:cartoon}). The interorbital interaction between $p_x$ and $p_y$ fermions is then turned on, which couples the motion of particles in the $x$ direction and that in the $y$ direction. In general, weakly coupled 1D Fermi chains with intrachain interaction cannot be studied by RPA because of the Luttinger liquid behaviors in such quasi-one-dimensional systems. The well-defined (fermionic) single-particle excitations, which are required by RPA, are absent in Luttinger liquids, which makes RPA invalid in such weakly coupled 1D Fermi chains~\cite{PhysRevLett.87.276405}. However, in our system, the key difference is the existence of the interorbital interaction that couples  $p_x$ and $p_y$ fermions, which makes our system two dimensional (2D). This can be understood as follows. (1) An elastic-scattering process in 1D between two particles with equal mass cannot change the momentum distribution of the two particles. Therefore, in a 1D chain or weakly coupled 1D chains with intrachain interaction, the momenta distribution $N(\mathbf{k}) \equiv \langle C^\dagger(\mathbf{k}) C(\mathbf{k}) \rangle$ cannot be changed by the elastic-scattering process, which causes the excitations to be collective and the system to be a Luttinger liquid, for which the RPA is invalid. In contrast, in our system, the interorbital scattering process is a 2D scattering process, which can change the momentum and energy of the $p_x$ fermion in $x$ chains by transferring some momentum and energy to the $p_y$ fermion in $y$ chains during the elastic-scattering process, and vice versa. In other words, the momentum distribution $N_{x} (\mathbf{k}) \equiv \langle C_{x}^\dagger(\mathbf{k}) C_{x}(\mathbf{k}) \rangle$ of the $p_x$ fermions can be changed by scattering, and the same for the $p_y$ fermions. As a result, although our system appears to be composed of 1D chains of weak interchain tunneling, its dynamics is fundamentally 2D.  At high temperature, a 2D system is at the Fermi-liquid phase with well-defined (fermionic) single-particle excitations, which is essentially different from the Luttinger liquid phase in 1D or quasi-one-dimensional systems. Therefore, RPA is valid in our 2D system. (2) We can also classify the interactions according to their relevance in the renormalization-group (RG) flow to the Fermi surfaces~\cite{1992hep.10046, RevModPhys.66.129}.
The dominant marginal term in our system is
$g \sum_{\tbf{k}_1, \tbf{k}_2} C^\dag _x (\tbf{k}_1+\tbf{Q}) C_x (\tbf{k}_1) C_y ^\dag(\tbf{k}_2 -\tbf{Q}) C_y(\tbf{k}_2) $,
where $\tbf{k}_1$, $\tbf{k}_2$, $\tbf{k}_1 +\tbf{Q}$ and $\tbf{k}_2 -\tbf{Q}$ are all near the Fermi surfaces~\cite{RevModPhys.66.129}.
This dominant interspecies interaction also induces effective intraspecies interactions.
The induced intraspecies forward and backward scattering processes are also marginal, but
they are much weaker since they are higher-order processes.
With the effect of the intraspecies interaction on the interspecies interaction neglected,
the RG flow of the interspecies interaction diverges for temperature
$T  \lesssim t \exp (- \alpha_1 t/|g|)$ ($\alpha_1$ is some constant).
We emphasize here that this interaction describes scattering processes in 2D rather than
1D. For such a 2D system with a single dominant interaction term, the RPA is well
justified~\cite{RevModPhys.66.129}.

\section{Derivation of the Field Theory}
\label{sec:eftapp}

The coupling constants $j,r,u,$ and $v$ in Equation~\eqref{eq:lcgl1} of Sec.~\ref{sec:eft} can be derived from the microscopic model. The Fermi-Hubbard model given by Equation~\eqref{eq:ham1} can be written in path integral form, where the partition function is given by $e^{-S_F}$ and the effective action reads
\begin{eqnarray}
S_F&=& \int d\tau \sum_{\mathbf{r},\alpha} \psi^\ast_\alpha(\mathbf{r},\tau) (\partial_\tau -\mu)\psi_\alpha(\mathbf{r},\tau)\nonumber\\
&& +\sum_{\mathbf{r} \alpha \beta}t_{\alpha \beta}
(\psi^{\ast}_{\alpha}(\mathbf{r}+\mathbf{e_{\beta}},\tau)
\psi_{\alpha}(\mathbf{r},\tau)+h.c.)\nonumber\\
&&+g\sum_{\mathbf{r}} \psi^\ast_x(\mathbf{r},\tau) \psi^\ast_y(\mathbf{r},\tau) \psi_y(\mathbf{r},\tau)\psi_x(\mathbf{r},\tau).
\label{eq:ft1}
\end{eqnarray}
In Equation~\eqref{eq:ft1}, the interaction term can be rewritten as
\begin{eqnarray}
\psi^\ast_x \psi^\ast_y \psi_y\psi_x=
\frac{(\psi^\ast_x \psi_x +\psi^\ast_y \psi_y)^2-(\psi^\ast_x \psi_x-\psi^\ast_y \psi_y)^2}{4}.\nonumber\\
\label{eq:ft2}
\end{eqnarray}
Consider two auxiliary Hubbard-Stratonovich fields $\int D (\rho_{1,2}) e^{S_{\rho_{1,2}}}$, where $S_{\rho_{1,2}}=\int d\tau \frac{g}{4} \sum_{\mathbf{r}} \rho_{1,2}^2(\mathbf{r},\tau)$. By shifting $\rho_{1,2} \rightarrow \rho_{1,2}-(\psi^\ast_x \psi_x \pm \psi^\ast_y \psi_y)$, $\rho_1$ and $\rho_2$ denote the total density field and density difference field, respectively. Multiplying  $\int D (\rho_{1,2}) e^{S_{\rho_{1,2}}}$ with shifted $\rho_{1,2}$ to $e^{-S_F}$, the quartic interaction between fermions in Equation~\eqref{eq:ft2} is eliminated. According to the mean-field analysis of CDW, the density difference has the saddle point (thermal average) at zero, which means we can ignore the density difference field $\rho_2$, since the fluctuation around zero is trivial. As a result, the interaction term in Eq.~\eqref{eq:ft1} is replaced by
\begin{equation}
-\frac{g}{4}\int d\tau \sum_{\mathbf{r}} \rho_1^2-2\rho_1(\psi_x^\ast \psi_x + \psi_y^\ast \psi_y).
\end{equation}
From mean-field analysis, the total density $\rho_1$ is fluctuating around momenta $0, \pm \mathbf{Q_1}$, and $\pm \mathbf{Q_2}$. The fluctuation around zero momentum is background density fluctuation, which is trivial. By ignoring such contribution, $\rho_1$ reduces to  $ \delta \rho$, which reproduces the total density fluctuation $\delta \rho$ around $\mathbf{Q_{1,2}}$ in Sec.~\ref{sec:eft}. In the long-wavelength limit, the density fluctuations around $\mathbf{Q_{1,2}}$ can be rewritten as
\begin{eqnarray}
\delta \rho(\mathbf{r},\tau)&=&\frac{T}{N^2} \sum_{|\mathbf{q}|<\Lambda, \omega,\sigma} [ \delta \rho_\sigma(\mathbf{Q_\sigma+q},\omega)e^{i\mathbf{(Q_{\sigma}+q)\cdot r}}e^{-i\omega \tau}\nonumber\\
&&+c.c.],
\label{eq:deltarho1}
\end{eqnarray}
by Fourier transform, where $\Lambda$ is some momentum cutoff of the long-wavelength limit, and $\sigma=1,2$. Recall that in Sec.~\ref{sec:eft}, the $\phi$ fields are defined through
\begin{equation}
\delta \rho (\mathbf{r},\tau)=\sum_{\sigma}\left [ \phi_{\sigma}(\mathbf{r},\tau) e^{i\mathbf{Q_{\sigma} \cdot r}} +c.c.\right].
\label{eq:deltarho2}
\end{equation}
By Fourier transform,
\begin{equation}
\textstyle{\phi_{\sigma}(\mathbf{r},\tau)=\frac{T}{N^2} \sum_{|\mathbf{q}|<\Lambda, \omega} e^{i(\mathbf{q\cdot r}-\omega\tau)}\phi_{\sigma} (\mathbf{q},\omega),}
\end{equation}
and comparing Equation~\eqref{eq:deltarho1} with Equation~\eqref{eq:deltarho2}, we reach the relationship $\delta \rho(\mathbf{Q_{\sigma}+q},\omega)=\phi_\sigma(\mathbf{q},\omega)$. The effective action is then written in momentum space, where $\delta \rho_\sigma$ can be replaced by $\phi_\sigma$. At last, the fermionic fields are integrated out and the $\phi$ fields are kept up to quartic terms, and we reach the expression
\begin{eqnarray}
S&=&\textstyle{\frac{1}{T}\int d^2 r~\sum_{\sigma=1,2} \left(j|\nabla \phi
_\sigma|^2+r|\phi_\sigma|^2+u|\phi_\sigma|^4\right)}\nonumber\\
&&\textstyle{+v|\phi_1|^2|\phi_2|^2 + S_t + ...,}
\label{eq:lcgl2}
\end{eqnarray}
which reproduces the field theory given by Equation~\eqref{eq:lcgl1} in Sec.~\ref{sec:eft}. The coefficients in Equation~\eqref{eq:lcgl2} are
\begin{eqnarray}
r&=&-\frac{g}{2T}-\frac{g^2}{4N^2T}\sum_{\mathbf{k}}\frac{1-2n_F(\xi_{\mathbf{k}})} {2\xi_{\mathbf{k}}},\nonumber\\
j&=&\frac{g^2}{16N^2T}\sum_{\mathbf{k}}\frac{\partial^2 n_F}{\partial\xi_{\mathbf{k}}^2} \frac{(t_\parallel\sin k_x)^2}{\xi_{\mathbf{k}}},\nonumber\\
u&=&\frac{g^4}{32N^2 T} \sum_{\mathbf{k}} \left( \frac{1-2n_F}{4\xi_{\mathbf{k}}^3}+\frac{\partial n_F}{ \partial \xi_{\mathbf{k}}} \frac{1}{2\xi_{\mathbf{k}}^2}\right),\nonumber\\
v&=&4u,
\label{eq:coefficient}
\end{eqnarray}
in static limit (we do not consider quantum fluctuations in the present work). Here, $\xi_{\mathbf{k}}$ is the spectrum of free $p_x$-orbital fermions.

In Sec.~\ref{sec:eft}, a $1/5$ commensurate filling case is considered. In order to obtain the coefficients in Eq.~\eqref{eq:coefficient} in this case by the above procedure, the chemical potential is adjusted to make the filling $1/5$ ($k_F=4/5 \pi$), with CDW momentum $\mathbf{Q}=(\frac{8\pi}{5},\frac{8\pi}{5})$.  A term $w\cos(5 \varphi)$ produced by $(\phi_\alpha)^5+c.c.$ also exists. $w$ in general is small because this term arises from high-order diagrams and is suppressed at finite temperature. The exact value of $w$ is not important, but the existence of this term is crucial in the commensurate filling. The field theory then reproduces the Z$_p$ compact clock model, and the RG analysis of this model can be used to determine the phase-transition temperature, as discussed in Sec.~\ref{sec:eft}.

\bibliography{pspinless}

\begin{thebibliography}{33}
\bibitem[{foo({\natexlab{b}})}]{footnote2}
\bibinfo{note}{This checkerboard order from interplay of $(2k_F,\pi)$ and
  $(\pi,2k_F)$ orders is different from the previous $(\pi,\pi)$ checkerboard
  order.}
\end{thebibliography}
\bibliographystyle{apsrev}

\end{document}